\documentclass[preprint,amsmath,eqsecnum,nofootinbib,tightenlines]{revtex4}
\usepackage{graphicx}

\begin{document}
\preprint{\begin{minipage}{24mm}{\small%
DESY 08-050\\%
KUNS-2138}
\end{minipage}}
\title{\vspace*{7mm}%
Low-scale Gaugino Mass Unification\vspace*{4mm}}
\author{Motoi Endo$^a$ and Koichi Yoshioka$^b$}
\affiliation{%
$^a$Deutsches Elektronen Synchrotron DESY, Notkestrasse 85, 22607
Hamburg, Germany\\
$^b$Department of Physics, Kyoto University, Kyoto 606-8502, Japan%
\vspace*{15mm}}
\begin{abstract}\vspace*{5mm}\noindent%
We study a new class of scenarios with the gaugino mass unification at
the weak scale. The general condition is first derived for the
unification to occur. Among the general cases, a particular attention
is drawn to the mirage gauge mediation where the low-energy mass
spectrum is governed by the mirage of unified gauge coupling which is
seen by low-energy observers. The gaugino masses have natural and
stable low-scale unification. The mass parameters of scalar quarks and
leptons are given by gauge couplings but exhibit no large mass 
hierarchy. They are non-universal even when mediated at the gauge
coupling unification scale. In addition, the gravitino is rather heavy
and not the lightest superparticle. These facts are in contrast to
existing gauge and mirage mediation models. We also present several
explicit models for dynamically realizing the TeV-scale unification.
\end{abstract}
\maketitle
\newpage

{\centering\section{Introduction}}

Supersymmetry is one of the most attractive frameworks for exploring
theoretical and phenomenological aspects of possible extensions of the
standard model (SM)~\cite{SUSYreview}. Supersymmetry is expected to be
broken around the electroweak scale. That is strongly suggested by the
supersymmetric unification of the SM gauge coupling constants. The
gauge coupling unification is obtained from the precise measurement of
coupling constants in the low-energy regime~\cite{GUT} and then probes
the existence of unification hypothesis in high-energy fundamental
theory.

The search for supersymmetry will be performed in various near future
experiments, e.g.\ through the dark matter probe of the lightest
superparticle. A more direct consequence of low-energy supersymmetry
is the observation of superpartners of the SM particles in the
forthcoming Large Hadron Collider, and the most important observable
quantity is the mass spectrum of new particles. The masses of
superparticles are generally expressed by soft breaking terms which do
not introduce quadratic divergences~\cite{soft}. These soft breaking
terms consist of gaugino masses, scalar masses, and scalar trilinear
couplings. They are induced by supersymmetry-breaking dynamics in
high-energy fundamental theory and are forced to have some special
properties in order to satisfy low-energy experimental constraints,
e.g.\ from flavor-changing rare processes~\cite{FCNC} and CP
violation~\cite{CP}. To this end, various scenarios of supersymmetry
breaking have been proposed in the literature, and each of them
predicts individual and distinctive signatures which would be observed
in coming experiments.

In this paper, we explore a new type of low-energy mass spectrum of
superparticles, where the spectrum around the electroweak scale is
predictive and directly written down in terms of high-energy
quantities. Under the hypothesis of gauge coupling unification in
fundamental theory, gaugino masses are unified in low-energy
regime. Furthermore, scalar quarks and leptons have no mass hierarchy
among them and also their masses are comparable in size to those of
gauginos. The situation that the low-energy spectrum tends to be
degenerate is similar to the supersymmetry-breaking
model~\cite{CJO,EYY} related to the moduli stabilization in string
theory. However our approach in this paper is the field-theoretical
construction of supersymmetry breaking and its mediation which induces
such a type of mass spectrum. The scenario is basically the gauge
mediated supersymmetry breaking~\cite{GM,GM2} in which the threshold
of messenger fields is affected by the super-Weyl anomaly
mediation~\cite{AM_RS,AM_GLMR} in supergravity. Therefore it shares
the phenomenological virtues with the gauge mediation, for example,
the suppression of serious higher-dimensional operators including
supersymmetry-breaking fields. The mass spectrum is, however, rather
different from existing scenarios and induces distinctive
phenomenology in particle experiments and cosmology. In particular,
the spectrum of scalar quarks and leptons is determined by gauge
charges and not universal even when they are mediated at the gauge 
coupling unification scale. The low-energy gaugino mass unification is
unaffected by changing the supersymmetry-breaking scale and also by
the existence of multiple thresholds. These facts are in contrast to
the model in Refs.~\cite{CJO,EYY}. On the other hand, the gravitino is
heavy and not the lightest superparticle, which is different from
typical gauge mediation models.

This paper is organized as follows. In Section~\ref{sec:MM}, we first
study the low-energy unification of gaugino masses in the simplest
case with the universality assumption. The general form of gaugino
masses with the low-energy unification is derived in
Section~\ref{sec:GMU} and its pattern is briefly classified in
Section~\ref{sec:class}. We also discuss in Section~\ref{sec:multi}
the low-energy unification in the presence of multiple threshold
scales of messenger fields and in particular examine whether the
unification scale is destabilized or not. Section~\ref{sec:scalar}
contains the formulas for supersymmetry-breaking parameters of
scalar fields. From Section~\ref{sec:MGM}, we focus on the gauge
mediation scenario with low-scale gaugino mass unification. In 
Section~\ref{sec:MGM}, we derive the general formula of mass spectrum
and discuss phenomenological aspects of the scenario. In
Section~\ref{sec:TeV}, the unification scale is supposed to be a TeV
scale, and various dynamical realization of TeV-scale unification are 
investigated including the effect of uplifting the vacuum energy. The
last section is devoted to summarizing the results and some discussion
of phenomenology.

\bigskip\bigskip\bigskip
{\centering\section{Mirage Mediation}
\label{sec:MM}}

Throughout this paper, $M_X$ denotes the scale at which we have soft
supersymmetry-breaking parameters generated by some high-energy
dynamics, e.g.\ from supergravity interactions, strongly-coupled gauge
sector, etc. In this section, we focus on the property of gaugino 
mass $M_\lambda$. Its general form at the scale $M_X$ is parameterized
as follows:
\begin{eqnarray}
  M_\lambda(M_X) \;=\; M_\lambda^X+\frac{bg^2(M_X)}{16\pi^2}F_\phi,
  \label{general}
\end{eqnarray}
where $g$ and $b$ are the gauge coupling constant and the one-loop
beta function coefficient of the corresponding gauge 
theory: $dg/d\ln\mu=bg^3/16\pi^2$ where $\mu$ is the renormalization
scale. The first term in \eqref{general} is the above-mentioned
supersymmetry-breaking effect from high-energy dynamics. It is given 
at tree or loop level of coupling constants in the theory and
generally depends on the energy 
scale: $M_\lambda^X=M_\lambda^X(M_X)$. The second term is called the
anomaly mediation effect~\cite{AM_RS,AM_GLMR} and comes from the
one-loop contribution of super-Weyl anomaly in supergravity. The 
latter effect generally exists in any supersymmetry-breaking theory
and must be taken into account. The $F_\phi$ contribution is usefully
expressed in terms of the compensator chiral multiplet $\Phi$ in the
conformal supergravity~\cite{SCsugra} and its value is given by fixing
the superconformal gauge transformation such 
that $\Phi=1+F_\phi\theta^2$ in the conformal frame.

The one-loop renormalization group for the gaugino mass below the 
scale $M_X$ is evolved down to the low-energy regime as
\begin{eqnarray}
  M_\lambda(\mu) \;&=&\; M_\lambda(M_X)\frac{g^2(\mu)}{g^2(M_X)}
  \nonumber \\
  &=&\; M_\lambda^X\left[\,1+\frac{bg^2(\mu)}{16\pi^2}
    \ln\left(\frac{\mu^2}{M_X^2}\right)+\frac{bg^2(\mu)}{16\pi^2}
    \frac{F_\phi}{M_\lambda^X} \right].
  \label{MRGE}
\end{eqnarray}
Here an important scale $\mu_m$ is introduced at which the running
effect [the second term in \eqref{MRGE}] and the anomaly mediation
effect [the third term in \eqref{MRGE}] are cancelled out. Note that
the complex phases of two contributions in \eqref{general} must be
aligned in order to have a real-valued $\mu_m$. From \eqref{MRGE}, we
obtain
\begin{eqnarray}
  \mu_m \;=\; M_X\exp\left(-F_\phi/2M_\lambda^X\right),
  \label{mm_scale}
\end{eqnarray}
and the gaugino mass at this scale reads
\begin{eqnarray}
  M_\lambda(\mu_m) \;=\; M_\lambda^X.
  \label{mm_gaugino}
\end{eqnarray}
The scale $\mu_m$ is determined only by the ratio of two
SUSY-breaking effects, and the gaugino mass at $\mu_m$ is given by the
contribution of high-energy dynamics, exclusive of the anomaly
mediation. It is interesting that we directly observe in low-energy
particle experiments the high-energy effect of supersymmetry breaking
in fundamental theory without being disturbed by ambiguous
renormalization-group effects.

As in supersymmetric extensions of SM, there are generally several
numbers of gauge groups in a theory. The mirage mediation, the
unification of gaugino masses (more generally, of superparticle
masses) at a low scale, is derived by the assertion that the 
scale $\mu_m$ can be defined independently of the gauge groups
considered. This condition is found from \eqref{mm_scale} to require
that $M_\lambda^X$'s are universal:
\begin{eqnarray}
  M_{\lambda_a}^X \;=\; (a \text{--independent}).
  \label{mm_cond}
\end{eqnarray}
That is, the mirage mediation scale $\mu_m$ can be obtained in the
case that high-scale dynamics generates the universal boundary value
for different gauginos. The condition does not need any details 
of $M_\lambda^X$. Furthermore, if the universality \eqref{mm_cond} is
realized, Eq.~\eqref{mm_gaugino} means that gaugino mass parameters at
the mirage scale, $M_{\lambda_a}(\mu_m)$, take the unified 
value ($=M_\lambda^X$). Consequently, the mirage mediation is found to
imply the mirage unification (of gaugino masses). The gauge coupling
unification is not necessarily needed and the only assumption is to
have the universal gaugino masses from some high-energy dynamics. For
example, the universal contribution comes from moduli fields in
supergravity or string theory. In particular, a recent scenario of
string-theory moduli stabilization~\cite{KKLT} is known to predict a
suppressed value of $M_\lambda^X$ relative to $F_\phi$ and then a
hierarchically (exponentially) small scale $\mu_m$ can be naturally
realized [see Eq.~\eqref{mm_scale}], which provides a characteristic
framework for low-energy phenomenology~\cite{MM_ph}.

It seems however that the universality \eqref{mm_cond} is only a
sufficient condition for the mirage mediation, where high-scale
effects directly appear as if by projected mirage in low-energy
regime. In what follows, we investigate more general situations for
the mirage unification to occur.

\bigskip\bigskip\bigskip
{\centering\section{General Mirage Unification}
\label{sec:GMU}}

The non-universality of superparticle spectrum is often generated by
high-energy physics at the mediation scale $M_X$\@. In this case, it
is a non-trivial issue to study what conditions are implied by
asserting the mirage unification (of gaugino masses) at a low-energy 
scale. From the discussion in the previous section, it is a naive
expectation that the mirage unification takes place if gaugino masses
from high-energy physics ``unify'' at some scale (except for the
anomaly mediation effect). Notice however that this unification scale
is virtual and not the mediation scale $M_X$.

Let us consider the simplest situation that there is one threshold of
supersymmetry-breaking dynamics at $M_X$\@. The general unification of
gaugino masses, $M_{\lambda_a}=M_{\lambda_b}$, is derived from the
low-energy renormalization-group evolution:
\begin{eqnarray}
  \left[ M_{\lambda_a}^X +\frac{b_ag_a^2(M_X)}{16\pi^2}F_\phi\right]
  \frac{g_a^2(\bar\mu_m)}{g_a^2(M_X)} \;=\;
  \left[ M_{\lambda_b}^X +\frac{b_bg_b^2(M_X)}{16\pi^2}F_\phi\right]
  \frac{g_b^2(\bar\mu_m)}{g_b^2(M_X)},
\end{eqnarray}
where $\bar\mu_m$ is the gaugino mass unification scale and the gauge
coupling constants at this scale are given by
\begin{eqnarray}
  \quad \frac{1}{g_x^2(\mu_m)} \;=\; \frac{1}{g_x^2(M_X)} 
  +\frac{b_x}{16\pi^2}\ln\left(\frac{M_X^2}{\mu_m^2}\right)
  \qquad (x=a,b).
\end{eqnarray}
Inserting these values into the above unification equation, we find
the unification scale
\begin{eqnarray}
  \bar\mu_m \;=\; M_X\exp\left[\frac{16\pi^2\big(M_{\lambda_a}^X
    -M_{\lambda_b}^X\big) +\big[b_ag_a^2(M_X)-b_bg_b^2(M_X)\big]
    F_\phi}{2b_bg_b^2(M_X)M_{\lambda_a}^X 
    -2b_ag_a^2(M_X)M_{\lambda_b}^X}\right],
  \label{mum}
\end{eqnarray}
and the unified value of gaugino masses
\begin{eqnarray}
  M_{\lambda_a}(\bar\mu_m) \;=\; M_{\lambda_b}(\bar\mu_m)
  \;=\; \frac{b_ag_a^2(M_X)M_{\lambda_b}^X
    -b_bg_b^2(M_X)M_{\lambda_a}^X}{b_ag_a^2(M_X)-b_bg_b^2(M_X)}.
  \label{gaumass}
\end{eqnarray}
It may be interesting to notice that the anomaly mediation effects are
dropped out in the expression of unified gaugino mass, though any
cancellation is not assumed between these and renormalization-group
effects.

We first study the unification condition that $\bar\mu_m$ is
independent of the gauge indices $a$ and $b$. If this condition is
satisfied, the scale $\bar\mu_m$ is entitled to the mirage unification
scale at which more than two gaugino masses take a common value. A
trivial solution is the universal 
contribution $M_{\lambda_a}^X=M_{\lambda_b}^X$ at the threshold
scale. In this case, the expressions \eqref{mum} and \eqref{gaumass}
become equivalent to the result of the mirage mediation discussed in
the previous section. To look for a more general solution, we rewrite
the bracket in the right-handed side of \eqref{mum} as
\begin{eqnarray}
  \frac{\frac{16\pi^2}{b_ab_bg_a^2(M_X)g_b^2(M_X)}
    \big(M_{\lambda_a}^X-M_{\lambda_b}^X\big)
    -\big[\frac{1}{b_ag_a^2(M_X)}-\frac{1}{b_bg_b^2(M_X)}\big]
    F_\phi}{M_{\lambda_a}^X/b_ag_a^2(M_X)
    -M_{\lambda_b}^X/b_bg_b^2(M_X)}.
  \label{mum2}
\end{eqnarray}
Since the supergravity interactions are universal, the coefficient 
of $F_\phi$ must be $a$,$b$-independent. The renormalization-group
running above the threshold scale are generally given by
\begin{eqnarray}
  \quad \frac{1}{g_x^2(M_X)} \;=\; \frac{1}{g_U^2} 
  +\frac{b_x+N_x}{16\pi^2}\ln\left(\frac{M_G^2}{M_X^2}\right)
  \qquad (x=a,b),
  \label{gRGE}
\end{eqnarray}
where $N_{a,b}$ denote the contribution of decoupled fields at the
threshold and $M_G$ is the scale at which the two running gauge
couplings $g_a$ and $g_b$ meet [$g_a(M_G)=g_b(M_G)\equiv g_U^{}$]. The
coefficient of $F_\phi$ in \eqref{mum2} now becomes
\begin{eqnarray}
  \frac{1}{b_ag_a^2(M_X)}-\frac{1}{b_bg_b^2(M_X)} \;=\;
  \Big(\frac{1}{b_a}-\frac{1}{b_b}\Big)\frac{1}{g_U^2}
  +\frac{1}{16\pi^2}\Big(\frac{N_a}{b_a}-\frac{N_b}{b_b}\Big)
  \ln\left(\frac{M_G^2}{M_X^2}\right).
\end{eqnarray}
Therefore we should have $N_a=N_b$ ($\equiv N$) as the first condition
for unification. With this condition at hand, we finally find that the
gauge factor independence of the exponent \eqref{mum2} leads to the
common value of gaugino masses:
\begin{eqnarray}
  M_{\lambda_a}^X(M_G) \;=\; M_{\lambda_b}^X(M_G),
\end{eqnarray}
with which not only the coefficient of $F_\phi$ but the first term 
of \eqref{mum2} becomes $a$,$b$-independent. Here the gaugino mass
factors $M_{\lambda_x}^X$ above the supersymmetry-breaking 
scale $M_X$ are virtually defined as if they obey the
renormalization-group equations corresponding to \eqref{gRGE}. Namely,
the second condition for unification to appear is that, at the scale
of gauge coupling unification, the corresponding gaugino masses also
(virtually) unify, except for the anomaly mediation effect. The form
of the unified value $M_{\lambda_x}^X(M_G)$ has no restriction and is
an arbitrary function of $g_U^{}$ and other universal couplings.

In summary, the general mirage unification (of gaugino masses) is
achieved in theory with gauge coupling unification and satisfies two
conditions: (i) the threshold of supersymmetry-breaking dynamics 
preserves the gauge coupling unification and (ii) gaugino masses from
supersymmetry-breaking dynamics virtually unify at the scale of gauge
coupling unification. As a result, the general form of gaugino masses
induced at the supersymmetry-breaking scale is consistent with the
low-energy mirage when it satisfies
\begin{eqnarray}
  M_\lambda^X \;=\; c_0+c_1g^2(M_X).
  \label{sol}
\end{eqnarray}
The coefficients $c_0$ and $c_1$ are universal (gauge-factor
independent) and, in particular, do not depend on $g(M_X)$. We have
also included the trivial solution (the $c_0$ term), corresponding to
the simple mirage mediation if $M_X=M_G$. The scale $\bar\mu_m$ of
the general mirage unification is written in a parallel fashion to the
previous case \eqref{mm_scale} as
\begin{eqnarray}
  \bar\mu_m \;=\; \bar M_X
  \exp\left(-F_\phi/2\bar M_\lambda^X\right).
  \label{gmu_scale}
\end{eqnarray}
The effective threshold scale $\bar M_X$ and the
supersymmetry-breaking mass parameter $\bar M_\lambda^X$ are defined as 
\begin{gather}
  \bar M_X \;\equiv\; M_G\left(\frac{M_X}{M_G}
  \right)^{c_0/\bar M_\lambda^X}\!, \qquad\;
  \bar M_\lambda^X \;\equiv\; c_0+c_1g_G^2,  \\[2mm]
  \frac{1}{g_G^2} \;\equiv\; \frac{1}{g_U^2}
  +\frac{N}{16\pi^2}\ln\left(\frac{M_G^2}{M_X^2}\right).
\end{gather}
These three quantities, $\bar M_X$, $\bar M_\lambda^X$, $g_G^{}$ are
found not to have gauge-group dependences and so $\bar\mu_m$ does
denote the low-energy unification scale of gaugino 
masses. From \eqref{gaumass} and \eqref{sol}, we evaluate the unified
value of gaugino masses at this scale:
\begin{eqnarray}
  M_{\lambda_a}(\bar\mu_m) \;=\; 
  M_{\lambda_b}(\bar\mu_m) \;=\;
  c_0+c_1g_G^2\,,
\end{eqnarray}
which is just equivalent to the effective boundary 
mass $\bar M_\lambda^X$. This fact implies that the effect of
high-energy physics is directly observed in low-energy regime as the
projection of mirage. It is also found, compared to \eqref{sol}, that
the low-energy value $M_\lambda(\bar\mu_m)$ is equal to the
dynamically induced mass $M_\lambda^X$ by replacing the gauge
coupling with $g_G^{}$. The ``coupling constant'' $g_G^{}$ represents
the effect of decoupled fields and naively seems to depend on the
threshold scale $M_X$\@. However we can show from the running
equations of gauge couplings that low-energy values of gauge couplings
are related to $g_G^{}$ as
\begin{eqnarray}
  \quad \frac{1}{g_G^2} \;=\; \frac{1}{g_x^2(\mu)}
  +\frac{b_x}{16\pi^2}\ln\left(\frac{\mu^2}{M_G^2}\right)
  \qquad (x=a,b),
  \label{grun}
\end{eqnarray}
This equation indicates an important property that $g_G^{}$ is
interpreted as the (virtual) unified value of gauge couplings in the
absence of any threshold and is determined only by (experimentally)
observed values of $g_x$ in low-energy regime. In 
particular, $g_G^{}$ does not depend on $M_X$ and therefore, the
low-energy unified value of gaugino 
masses $M_{\lambda_x}(\bar\mu_m)$ is also insensitive to the threshold
scale. Further it is interesting to notice that, similarly to the
gaugino masses, the gauge coupling constants are also mediated by
mirage from high to low-scale physics: in future particle experiments,
we would directly probe the high-energy unified value $g_G^{}$ through
the determination of superparticle masses. Probing high-energy physics
without being disturbed by intermediate-scale unknown factors will
clarify the mechanism of supersymmetry breaking as well as grand
unified theory.

As we have shown, the general mirage unification can be defined even
when the spectrum is non-universal at the supersymmetry-breaking scale
and the coupling unification scale. The general formulas are given 
by \eqref{sol} and \eqref{gmu_scale} with the mirage value of unified
gauge coupling which is evaluated only by low-energy observables and
independent of supersymmetry-breaking thresholds. One remark is that
the unification scale $\bar\mu_m$ does not make sense unless the
complex phases of $c_0$ and $c_1$ terms are aligned to that of the
anomaly mediation $F_\phi$. That may restrict possible dynamics of
supersymmetry breaking and its mediation.

\bigskip\bigskip\bigskip
{\centering\section{Classification}
\label{sec:class}}

The gaugino masses in the scenario of general mirage unification have
the form \eqref{sol}. We briefly discuss each case separately and
comment on possible dynamics of supersymmetry-breaking mediation
sector.

\bigskip\bigskip
{\centering\subsection{$c_0\neq0$, $\;c_1=0$}}

The first simple case is that the $c_0$ term is dominant. In this
case, gaugino masses are universal at the supersymmetry-breaking 
scale $M_X$\@. The universal contribution originates from, 
e.g., gravitational interactions, moduli fields in high-energy theory,
and so on. That results in the simple mirage mediation discussed in
Section~\ref{sec:MM}.

\bigskip\bigskip
{\centering\subsection{$c_0=0$, $\;c_1\neq0$}}

The second case is that the gauge threshold contribution is 
dominant: $c_0=0$ and $c_1\neq0$. That is understood as the situation
that supersymmetry breaking is mediated by some gauge interactions at
loop level. In this case, $M_\lambda^X$'s become universal if they
were interpolated to the gauge coupling unification scale, and then
the discussion returns to the simple mirage mediation with the
threshold scale $M_X=M_G$\@. However there is one difference that the
gauge coupling in the interpolated mass $M_\lambda^X(M_G)$ is not the
real value $g_U^{}$ but the virtual one $g_G^{}$, which represents the
deviation due to the presence of supersymmetry-breaking dynamics 
above $M_X$\@. In other words, if one uses the mirage scale 
formula \eqref{mm_scale} with the interpolated 
mass $M_\lambda^X(M_G)$, the threshold scale should be modified
accordingly.

As an explicit example, let us see the following form of gaugino masses:
\begin{eqnarray}
  M_{\lambda_a}(M_X) \;=\; \frac{g_a^2(M_X)}{16\pi^2}F+
  \frac{b_ag_a^2(M_X)}{16\pi^2}F_\phi\,.
\end{eqnarray}
The superparticle mass spectrum is given by the sum of the gauge and
super-Weyl anomaly contributions. The relative complex phase of 
two $F$ terms, $F$ and $F_\Phi$, should be aligned from a
phenomenological analysis of CP violation~\cite{EYY2}. A simple
dynamical example is the so-called deflected anomaly
mediation~\cite{DAM}. The above expression 
means $c_0=0$ and $c_1=F_X/16\pi^2$, and hence the mirage unification
scale is found
\begin{eqnarray}
  \bar\mu_m \;=\; M_G\,\exp\left(\frac{-8\pi^2}{g_G^2}
    \frac{F_\phi}{F}\right),
\end{eqnarray}
where the mirage value of unified gauge coupling $g_G^{}$ is
determined by the observed values of gauge couplings at a low-energy
scale $\mu$:
\begin{eqnarray}
  \frac{1}{g_G^2} \;=\; \frac{1}{g_x^2(\mu)}
  +\frac{b_x}{16\pi^2}\ln\left(\frac{\mu^2}{M_G^2}\right).
\end{eqnarray}

\bigskip\bigskip
{\centering\subsection{$c_0\neq0$, $\;c_1\neq0$}}

The last one is the most general case and normally needs two sources
of supersymmetry breaking. A simple example is the coexistence of the
contributions via supergravity and gauge 
interactions~\cite{GG,XT} from several supersymmetry-breaking
sectors. It is also possible to realize this type of spectrum with a
single source of supersymmetry breaking. For this purpose, let us
assume the following schematic Lagrangian:
\begin{eqnarray}
  \int\!d^2\theta\,\bigg[
    \Big(\frac{1}{4g^2}+\frac{X}{M_{\rm pl}}\Big)W^\alpha W_\alpha 
    +\big(M_\Psi+X\big)\bar\Psi\Psi\bigg]+{\rm h.c.} \,+
    \text{(dynamics for $X$)},
\end{eqnarray}
where $X$ is the representative field of supersymmetry breaking which
has a non-vanishing $F$ component, and $\Psi$, $\bar\Psi$ are the
vector-like messenger multiplets. The first term gives a tree-level
gravity contribution to gaugino masses of the form 
of $F_X/M_{\rm pl}$. The second term induces a mass splitting in each
messenger multiplet and gives the gauge contribution from a one-loop
diagram involving the messenger fields. Thus the gauge contribution
takes the form of $(1/16\pi^2)(F_X/M_\Psi)$. If the messenger mass
scale $M_\Psi$ is smaller than the gravity scale $M_{\rm pl}$ by
one-loop order quantity, the two contributions of supersymmetry
breaking are comparable to each other and equally important for
phenomenology such as the modification of low-energy unification scale
and superparticle mass spectrum.

\bigskip\bigskip\bigskip
{\centering\section{Multi Thresholds and Stability of Mirage}
\label{sec:multi}}

In this section we study the case that there exist multiple threshold
scales of supersymmetry breaking dynamics. In addition to the 
scale $M_X$ previously discussed, superparticles are supposed to
receive the contribution of supersymmetry-breaking masses from
different dynamics at $M_X'$, which is assumed to be smaller 
than $M_X$ without the loss of generality. In particular, we examine
whether the mirage unification is spoiled or not in the presence of
additional thresholds.

\bigskip\bigskip
{\centering\subsection{Simple Mirage Case ~$(c_0\neq0$, $\;c_1=0)$}}

Let us first consider the simple mirage case ($c_1=0$) analyzed in
Section~\ref{sec:MM}. We have additional gaugino mass 
contribution $M_\lambda'$ at the scale $M_X'$\@. It is noted that the
net contribution at this threshold is the sum of $M_\lambda'$ and the
supersymmetric contribution which compensates the anomaly
mediation. In low-energy regime ($\mu<M_X'$), the gaugino mass is
given by the one-loop renormalization-group flow:
\begin{eqnarray}
  M_\lambda(\mu) &=& \Big[M_\lambda(M_X')
  +M_\lambda'\Big]\frac{g^2(\mu)}{g^2(M_X')} \nonumber \\[1mm]
  &=& \left(M_\lambda^X+M_\lambda'\right)
  \left[1+\frac{b'g^2(\mu)}{16\pi^2}
    \ln\left(\frac{\mu^2}{M_X'^{\,2}}\right)\right]  
  +\frac{b'g^2(\mu)}{16\pi^2}F_\phi+\frac{bg^2(\mu)}{16\pi^2}
  M_\lambda^X\ln\left(\frac{M_X'^{\,2}}{M_X^2}\right)\!,\qquad
\end{eqnarray}
where $b'$ is the beta function coefficient of gauge 
coupling $g$ below the threshold scale $M_X'$\@. Repeating the
previous analysis, the new scale of low-energy unification is formally
written down as
\begin{eqnarray}
  \bar\mu_m' \;=\; \bar\mu_m\,
  \bigg(\frac{M_X'}{M_X}\bigg)^{\frac{b'-b}{b'}
    \frac{M_\lambda^X}{M_\lambda^X+M_\lambda'}}
  \bigg(\frac{M_X'}{\bar\mu_m}\bigg)^{\frac{M_\lambda'}{M_\lambda^X
      +M_\lambda'}}\,.
\end{eqnarray}
The unification scale $\bar\mu_m$ in the single threshold case has
been defined in \eqref{mm_scale}. It is found from this expression
that, in order for $\bar\mu_m'$ to be the unification scale, the
following two conditions are additionally required: (i) the threshold
contribution $M_\lambda'$ is universal and (ii) the ratio of beta
functions $b/b'$ is independent of gauge groups. The latter condition
is rather restrictive. The general solution to the latter condition is
given by $b=b'$ which implies an unrealistic situation that decoupled
fields at either threshold are only gauge singlets. Moreover, one
notices that $\bar\mu_m'$ is no longer a mirage unification scale,
even if the threshold contribution is supersymmetric ($M_\lambda'=0$)
or grand unification like ($b_a-b_a'=\text{universal}$).

\bigskip\bigskip
{\centering\subsection{Gauge Threshold Case ~$(c_0=0$, $\;c_1\neq0)$}
\label{sec:c1}}

Another typical case has the contribution of gauge threshold 
only ($c_0=0$), i.e.\ the scenario with gauge and anomaly
mediated supersymmetry breaking. Let us consider an additional gauge
threshold at $M_X'$\@. Its form is written down 
as $M_\lambda'=c_1'g^2(M_X')$ where the coefficient $c_1'$ is
universal for different gaugino masses. In the low-energy 
regime ($\mu<M_X'$), the gaugino mass is given by the one-loop
renormalization-group flow:
\begin{eqnarray}
  M_\lambda(\mu) \;&=&\, \bigg[ \Big(c_1g^2(M_X)
  +\frac{bg^2(M_X)}{16\pi^2}F_\phi\Big)\frac{g^2(M_X')}{g^2(M_X)}
  +c_1'g^2(M_X') +\Delta_{\rm AM}(M_X')\bigg] 
  \frac{g^2(\mu)}{g^2(M_X')} \nonumber \\
  &=&\; (c_1+c_1')g^2(\mu)+\frac{b'g^2(\mu)}{16\pi^2}F_\phi,
\end{eqnarray}
where $b'$ is the beta function coefficient for gauge 
coupling $g$ below the threshold scale $M_X'$\@. The last 
quantity $\Delta_{\rm AM}$ denotes the supersymmetric threshold
correction which preserves the ultraviolet insensitivity of super-Weyl
anomaly mediation. It is found that, in the previous expressions for
the single threshold case, $c_1$ should be shifted to $c_1+c_1'$, and
further, $M_X$ and $b$ are replaced with $M_X'$ and $b'$. The last
point we should take into account is the modification of the
renormalization group running of gauge couplings. In the case of
multiple thresholds at $M_X$ and $M_X'$, the gauge couplings take the
unified value $g_U'$ at $M_G'\,$:
\begin{eqnarray}
  \quad \frac{1}{g_x^2(M_X')} \;=\; \frac{1}{g_U'^{\,2}}
  +\frac{b_x}{16\pi^2}\ln\left(\frac{M_X^2}{M_X'^2}\right)
  +\frac{b_x+N}{16\pi^2}\ln\left(\frac{M_G'^{\,2}}{M_X^2}\right)
  \qquad (x=a,b).
  \label{grun2}
\end{eqnarray}
Repeating the previous analysis of mirage unification with this
modified running equation, we find that the mirage unification is
preserved for grand unification like threshold, that 
is, $b_a-b_a'=b_b-b_b'$ ($\equiv N'$). At the same time, the gauge
coupling unification scale is not modified: $M_G=M_G'$. In the end,
the mirage unification scale in the multi threshold case is given by
\begin{eqnarray}
  \bar\mu_m' \;=\; M_G\,\exp\big(\!-F_\phi/2\bar M_\lambda^{X'}\big).
\end{eqnarray}
The effective boundary mass $\bar M_\lambda^{X'}$ is defined as
\begin{gather}
  \bar M_\lambda^{X'} \;\equiv\; (c_1+c_1')\,g_G'^{\,2}\,, \\[2mm]
  \frac{1}{g_G'^{\,2}} \;\equiv\; \frac{1}{g_U'^{\,2}}
  +\frac{N'}{16\pi^2}\ln\left(\frac{M_X^2}{M_X'^2}\right)
  +\frac{N+N'}{16\pi^2}\ln\left(\frac{M_G^2}{M_X^2}\right).
  \label{gbarp}
\end{gather}
Since $\bar M_\lambda^{X'}$ and $g_G'$ do not have gauge-group
dependences, the new scale $\bar\mu_m'$ is properly defined as the
mirage unification scale (of gaugino masses). The unified value of
gaugino masses at $\bar\mu_m'$ is evaluated as
\begin{eqnarray}
  M_{\lambda_a}(\bar\mu_m') \;=\; 
  M_{\lambda_b}(\bar\mu_m') \;=\; 
  (c_1+c_1')\,g_G'^{\,2}\,,  
\end{eqnarray}
which is equal to the effective boundary 
mass $\bar M_\lambda^{X'}$. Compared with the single threshold case,
the virtual coupling $g_G^{}$ seems to be modified to $g_G'$ due to 
the effect of the additional threshold. However we can 
show from \eqref{grun2} and \eqref{gbarp} that low-energy values of
gauge couplings ($\mu<M'$) become
\begin{eqnarray}
  \quad \frac{1}{g_G'^{\,2}} \;=\; \frac{1}{g_x^2(\mu)}
  +\frac{b_x'}{16\pi^2}\ln\left(\frac{\mu^2}{M_G^2}\right)
  \qquad (x=a,b).
\end{eqnarray}
This equation indicates that $g_G'$ does not depend both 
on $M_X$ and $M_X'$, i.e.\ insensitive to the presence of
supersymmetry-breaking dynamics. It is also interesting to find that,
compared with the single threshold case \eqref{grun}, $g_G'$ is
equivalent to $g_G^{}$ for fixed low-energy observables, and so equal
to the (mirage) unified gauge coupling without any thresholds:
\begin{eqnarray}
  g_G^{} \;=\; g_G'\,.  
\end{eqnarray}
The real unified gauge coupling $g_U'$, of course, becomes 
different from $g_U^{}$ and sensitive to the presence of thresholds.

In summary, for the gauge threshold case, the mirage unification is
preserved even when there exist multiple thresholds of
supersymmetry-breaking dynamics. The mirage scale does not explicitly
depend on the threshold scales (the messenger mass scales). The only
influence of multiple thresholds is the cumulative effect 
of $c_1$ terms in gaugino mass. These facts show that only the total
number of messenger fields is relevant. Finally, the mirage
unification is not spoiled by supersymmetric threshold ($c_1'=0$),
unlike the simple mirage case.

\bigskip\bigskip\bigskip
{\centering\section{Supersymmetry Breaking Terms for Scalars}
\label{sec:scalar}}

We have discussed supersymmetry-breaking mass parameters for
gauginos. Scalar superparticles also receive similar effects from
their couplings to supersymmetry-breaking fields $X$ and $\Phi$. The
result is expressed in terms of soft mass parameters: trilinear and
bilinear holomorphic couplings and non-holomorphic scalar masses
squared. In this section, we present the general formulas for scalar
supersymmetry-breaking terms.

As seen above, the effect of super-Weyl anomaly is important in
discussing the gaugino mass unification and then, the compensator
formalism of supergravity is useful for deriving the general form of
supersymmetry-breaking terms for scalars. For scalar supermultiplets,
the supergravity Lagrangian is given by two ingredients, i.e.\ the
K\"ahler potential $K$ and superpotential $W$:
\begin{eqnarray}
  {\cal L} \;=\; \int\!d^4\theta\,\Phi^\dagger\Phi\,
  f(Q_i,Q_i{}^\dagger\!,X,X^\dagger\!,\Phi,\Phi^\dagger) \,+
  \Big[\int\!d^2\theta\,\Phi^3\,W(Q_i,X) \,+{\rm h.c.}\Big],
\end{eqnarray}
where $Q_i$ denote the scalar superfields for which we now want to
derive the supersymmetry-breaking terms. The supergravity $f$ function
is related to the K\"ahler potential as $f=-3e^{-K/3}$. We have taken
into account the fact that the K\"ahler potential has the
quantum-level dependence on the compensator field $\Phi$. Note that
the superpotential is known to be protected from radiative corrections
due to the non-renormalization theorem and have no $\Phi$
dependence. As in the case of gaugino masses, the $\Phi$-dependent
pieces come out through the renormalization procedure and induce the
anomaly-mediated contribution of supersymmetry breaking. When
including quantum effects, it may be easier to analyze the scalar
potential in the conformal frame of supergravity where the
superconformal gauge symmetry is fixed by 
choosing $\Phi=1+F_\phi\theta^2$. The supergravity Lagrangian in the
Einstein frame, where the (super)gravity kinetic terms are canonical,
is obtained by the specific super-Weyl transformation~\cite{Ein} and
the scalar potential analysis in this frame can be performed with the
gauge fixing condition~\cite{improved}: $\Phi=
e^{K/6}[1+(F_\phi+\frac{1}{3}K_iF_i)\theta^2]$\@. It is noted that
there is no difference between these two gauge choices for deriving
the leading order supersymmetry-breaking terms 
if $|K_iF_i|\ll|F_\phi|$\@. This condition is obviously satisfied 
when $|F_X/X|\sim|F_\phi|$ and $|X|\ll1$ as in the case of mirage
unification with gauge thresholds (and also satisfied in the case of
simple mirage mediation 
where $|F_X/X|\ll|F_\phi|$ and $|X|\sim1$). Therefore in the following
we study the scalar supersymmetry-breaking terms in the conformal
frame of supergravity.

To see the supersymmetry-breaking terms of scalar fields $Q_i$, we
first integrate out the auxiliary components $F_{Q_i}$ via their
equations of motion:
\begin{eqnarray}
  F_\phi^*f_{Q_i}^{}+W_{Q_i}+\sum_I F_I^\dagger f_{Q_iI^\dagger}
  \;=\; 0,
\end{eqnarray}
where the lower indices of $f$ and $W$ denote the field
derivatives. The index $I$ runs over all the chiral multiplet scalars
in the theory, i.e.\ $I=Q_i,X,\Phi$ in the present case. After the
integration, the resultant scalar potential is given by
\begin{eqnarray}
  V &=& \Big(\frac{f_Qf_{Q^\dagger}}{f_{QQ^\dagger}}-f\Big) 
  F_\phi^*F_\phi +\!\sum_{I,J\neq Q}
  \Big(\frac{f_{QI^\dagger}f_{JQ^\dagger}}{f_{QQ^\dagger}}
  -f_{JI^\dagger}\Big) F_I^\dagger F_J 
  +\!\bigg[\sum_{I\neq Q}
  \Big(\frac{f_Qf_{IQ^\dagger}}{f_{QQ^\dagger}}-f_I\Big) 
  F_\phi^*F_I +{\rm h.c.}\bigg]   \nonumber \\
  && \qquad +\Big[\,f_{QQ^\dagger}^{-1}W_Q
  \Big(\frac{1}{2}W_{Q^\dagger}^*+F_\phi f_{Q^\dagger} +
  \sum_{I\neq Q}F_If_{IQ^\dagger}\Big) 
  -3WF_\phi -W_XF_X+{\rm h.c.}\Big].
  \label{V}
\end{eqnarray}
We have dropped the flavor index $i$ of $Q_i$ just for notational
simplicity. Note that the derivative indices $I,J$ contain the
compensator field $\Phi$ which leads to radiative effects through the
supergravity anomaly. It easily turns out that the first line 
in \eqref{V} generates non-holomorphic scalar mass terms and the
second one holomorphic supersymmetry-breaking couplings as well as a
possible supersymmetric mass term contained in $|W_Q|^2$.

\bigskip\bigskip
{\centering\subsection{Holomorphic Scalar Couplings}}

The scalars $Q_i$ acquire supersymmetry-breaking holomorphic
couplings, including trilinear and bilinear ones in scalar fields
(usually called the $A$ and $B$ terms, respectively). They are induced
in the presence of corresponding superpotential terms in $W$. The most
general expression of holomorphic supersymmetry-breaking terms can be
calculated from the second line of the supergravity scalar 
potential \eqref{V}. For practical purposes, it is almost sufficient
to know holomorphic scalar couplings for the minimal K\"ahler 
form $K=Z_QQ^\dagger Q$ where the wavefunction factor depends 
on $\Phi$ through the 
renormalization: $Z_Q=Z_Q(X,X^\dagger,\Phi,\Phi^\dagger)$. In this
case we find that the second line in the potential \eqref{V} induces
\begin{eqnarray}
  {\cal L}_A \;=\; 
  F_X\bigg[W_X-\sum_Q\frac{\partial\ln Z_Q}{\partial X}
  \frac{\partial W}{\partial\ln Q}\bigg] +F_\phi\bigg[3W
  -\sum_Q\Big(1+\frac{\partial\ln Z_Q}{\partial \phi}\Big)
  \frac{\partial W}{\partial\ln Q}\bigg] +{\rm h.c.}.
\end{eqnarray}
The first term (the $F_XW_X$ term) is irrelevant unless the scalar
multiplets $Q_i$ directly couple to $X$ in the superpotential. As an
example, let us consider the superpotential with Yukawa and mass
terms; $W=y_{ijk}Q_iQ_jQ_k+\mu_{ij}Q_iQ_j$. The corresponding
trilinear and bilinear supersymmetry-breaking couplings are read off
from the general expression ${\cal L}_A$ and are given by
\begin{eqnarray}
  A_{ijk} \;&=&\; \sum_{I=X,\phi}
  \frac{\partial\ln(Z_{Q_i}Z_{Q_j}Z_{Q_k})}{\partial I}F_I, \\
  B_{ij} \;&=&\; -F_\phi +\sum_{I=X,\phi}
  \frac{\partial\ln(Z_{Q_i}Z_{Q_j})}{\partial I}F_I,
\end{eqnarray}
for the definition of Lagrangian 
parameters: ${\cal L}=-A_{ijk}y_{ijk}Q_iQ_jQ_k-B_{ij}\mu_{ij}Q_iQ_j
+{\rm h.c.}$. The $\phi$ derivative is translated to the energy-scale
dependence of wavefunction factors and the coefficients 
of $F_\phi$ are given by the anomalous dimensions of scalar fields. On
the other hand, the $X$ dependence of $Z$ is fixed model-dependently
and its supersymmetry-breaking effects have some variety.

We have two brief comments on the phenomenological aspect of these
formulas. First it is noted that the supersymmetry-breaking parameters
are described by $F_X/X$ and $F_\phi$ with real
coefficients. Therefore if the complex phases of these two $F$ terms 
are aligned, phases of supersymmetry-breaking parameters including
gaugino masses can be rotated away with one suitable $R$ symmetry
rotation, and the CP symmetry is not violated in the
supersymmetry-breaking sector. Second, the above $B$-term formula,
when applied to the minimal supersymmetric SM and beyond, causes a too
large value of the $B$ parameter to trigger the correct electroweak
symmetry breaking, if the $F_\phi$ contribution is dominant. While
there have been several proposed solutions to this
problem~\cite{AM_RS,DAM,Bmu}, they are model-dependent and generally
predict different values of $B$ according to how to develop $\mu$
parameters.

\bigskip\bigskip
{\centering\subsection{Non-holomorphic Scalar Masses}}

Scalar fields generally receive non-holomorphic supersymmetry-breaking
masses from their couplings to supersymmetry-breaking fields. The mass
spectrum of superpartners of quarks and leptons is sensitive to the
detailed form of K\"ahler potential which, in turn, is restricted by
phenomenological constraints. Here we suppose the minimal K\"ahler
potential $K=Z_QQ^\dagger Q$ as in the previous section. The 
possible $X$ dependence of the wavefunction factor is determined,
depending on the property of $X$, by claiming the absence of
flavor-changing higher-dimensional operators~\cite{EYY}. We do not
discuss further here and derive the general formula for
supersymmetry-breaking scalar masses.

The non-holomorphic mass terms come from the first line of the
potential \eqref{V}. Expanding about $Q$, we obtain the general
expression for the minimal K\"ahler form:
\begin{eqnarray}
  m_Q^2 \;=\; \sum_{I,J=X,\phi}
  \frac{\partial^2\ln Z_Q^{-1}}{\partial I^\dagger \partial J}
  F_I^\dagger F_J^{},
\end{eqnarray}
for the canonical normalization of the $Q_i$ field kinetic term. The
second-order derivative with respect to the compensator $\Phi$ leads
to the anomaly mediated contribution to supersymmetry-breaking
masses. A more essential ingredient is the cross term of 
two $F$-component effects $F_X$ and $F_\phi$\@. This part is found to
play an important role in discussing the mirage behavior of
superparticle masses.

\bigskip\bigskip\bigskip
{\centering\section{Mirage Gauge Mediation}
\label{sec:MGM}}

Among the general mirage unification scenarios classified in the
previous section, the gauge threshold case is shown to have natural
and stable low-energy unification against possible but obscure
intermediate thresholds. The gauge threshold scenario is also favored
from phenomenological viewpoints such as the suppression of rare
processes beyond the SM\@ and the cosmology. In the rest of this
paper, we focus on analyzing this class of scenario, which is called
here the mirage gauge mediation.

\bigskip\bigskip
{\centering\subsection{Setup and Supersymmetry Breaking Terms}}

We first study a simple gauge threshold model, and discuss its mirage
unification behavior and superparticle spectrum. Let us consider the
following form of Lagrangian:
\begin{eqnarray}
  {\cal L} \;&=& \int\!d^4\theta\,\Phi^\dagger\Phi\,
  Z_i(X,X^\dagger\!,\Phi,\Phi^\dagger) Q_i^\dagger Q_i^{}\,
  +\!\int\!d^2\theta\,S(X,\Phi)W^\alpha W_\alpha 
  \,+{\rm h.c.} \nonumber \\
  && \qquad\qquad +\!\int\!d^2\theta\,\Phi^3 X\bar\Psi\Psi\,+{\rm h.c.}
  +\text{(dynamics for $X$)},
  \label{LagMGM}
\end{eqnarray}
where $Q_i$ and $W^\alpha$ denote the matter and gauge chiral
superfields with the renormalization factors $Z_i$ and $S$,
respectively. The vector-like messenger 
multiplets $\Psi$ and $\bar\Psi$ belong to grand unification-like
representations, i.e.\ they give the universal contribution to the SM
gauge beta functions in order to preserve the gauge coupling
unification in the absence of threshold. The K\"ahler $f$ function is
expanded by $Q_i$ and only the leading kinetic term is included. The
leading constant ($Q_i$-independent) term was not explicitly written
but its significance will be discussed in later sections. In what
follows, we assume as an example that the messenger multiplets compose
of $N$ pairs of 5-plets and its conjugates of $SU(5)$\@. The
compensator superfield $\Phi$ controls the Weyl invariance of the
theory and its scalar and fermionic components are fixed by the
superconformal gauge transformation. Finally, $X$ is the
representative chiral superfield of supersymmetry breaking and its
expectation value is determined by high-energy dynamics of 
stabilizing $X$ such that $X=M_X+F_X\theta^2$.\footnote{Here $M_X$ is
a dimension-less parameter. The threshold mass scale is given by the
expectation value of the scalar component of dimension-one 
superfield $\tilde X\equiv X\Phi$. If the following formulas are
expressed in terms of $F_{\tilde X}$ instead of $F_X$, the
anomaly-mediated contribution should be replaced with the one
evaluated above the threshold scale.} The dynamics for $X$ field is
unspecified here and will be explicitly discussed with various
examples in supergravity towards constructing a fully viable theory.

The basic building blocks in \eqref{LagMGM} are parallel to the
deflected anomaly mediation, particularly the Pomarol-Rattazzi
model~\cite{DAM}, and its phenomenological aspects have been
discussed~\cite{DAM_ph}. We here use this as a simple model for 
illustrating the gauge threshold case. However our main focus is on
the low-scale unification, which does not necessarily mean the
deflected anomaly mediation. It is sufficient for the mirage gauge
mediation to have an independent source of gaugino masses proportional
to $g^2$. Various types of such source have been found in the
literature and they all belong to the class with $c_0=0$. In later
sections, we will be interested in discussing the property of mirage
unification at observable (TeV) scale, finding the necessary
conditions for realizing it, and presenting several dynamical
mechanisms to satisfy the conditions. For example, some new ingredient
is needed to lower the messenger scale and to properly modify 
the $F$-term ratio, and so we propose various dynamics [the second
line in \eqref{LagMGM}] and (hidden-sector) effects by performing
explicit construction of the extensions and detailed analysis.

The wavefunction factors $Z_i$ and the gauge kinetic 
function $S$ depend on the supersymmetry-breaking field $X$ at quantum
level. The tree-level $X$ dependences through higher-dimensional
operators are suppressed by the cutoff scale $M_{\rm pl}$ which is
much larger than the messenger scale $M_X$\@, otherwise these
operators sometimes induce disastrous phenomenology such as
flavor-changing rare processes and CP violations. This suppression is
known to be one of the virtues of gauge mediated supersymmetry
breaking and our present model shares this excellent property. The
compensator dependence also appears at loop level due to the classical
scale invariance. Therefore the supersymmetry-breaking effects are
extracted by turning on the $F$ components and by expanding the
quantum-level dependence~\cite{sse} with respect 
to $F_X$ and $F_\phi$. The renormalization factors in low-energy
region and their dependences on $X$ and $\Phi$ are obtained from the
solutions of one-loop renormalization-group equations in the
superfield forms:
\begin{eqnarray}
  Z_i(\mu) \;&=&\; Z_i(\Lambda)
  \left(\frac{\text{Re}\,S(X\Phi)}{\text{Re}\,S(\Lambda)}
  \right)^{\frac{2C_i}{b+N}}\!
  \left(\frac{\text{Re}\,S(\mu)}{\text{Re}\,S(X\Phi)}
  \right)^{\!\frac{2C_i}{b}}, \\[1mm]
  S(\mu) \;&=&\; S(\Lambda) +\frac{b+N}{32\pi^2}\ln
  \left(\frac{\Lambda}{X}\right)
  +\frac{b}{32\pi^2}\ln\left(\frac{X\Phi}{\mu}\right) \,,
\end{eqnarray}
where $b$ is the one-loop beta-function coefficient below the
threshold scale and $C_i$ denotes the quadratic Casimir, explicitly
given below. The scale $\Lambda$ means some high-energy initial 
point ($\Lambda>M_X$) above which no supersymmetry-breaking dynamics
exists.\footnote{We assume that the messenger 
fields $\Psi$ and $\bar\Psi$ do not have soft supersymmetry-breaking
parameters above $M_X$. If not so, the more general
formulas~\cite{MNY} should be utilized for deriving soft terms for
low-energy fields.}

The soft supersymmetry-breaking mass parameters, gaugino 
masses $M_{\lambda_a}$, scalar trilinear couplings $A_i$,
non-holomorphic scalar masses $m_i^2$, are then derived from the
general formulas given in the previous section:
\begin{eqnarray}
  M_{\lambda_a}(\mu) \;&=&\; 
  \frac{-Ng^2_a(\mu)}{16\pi^2}\frac{F_X}{M_X}
  +\frac{b_ag_a^2(\mu)}{16\pi^2}F_\phi,  \\[1mm]
  A_i(\mu) \;&=&\; \frac{NC^a_i}{8\pi^2b_a}
  \Big[ g^2_a(\mu)-g^2_a(M_X)\Big]\frac{F_X}{M_X}
  -\frac{C^a_ig^2_a(\mu)}{8\pi^2}F_\phi, \\[1mm]
  m^2_i(\mu) \;&=&\; \frac{NC^a_i}{128\pi^4b_a}\Big[ (b_a+N)g^4_a(M_X)
  -Ng^4_a(\mu)\Big]\bigg|\frac{F_X}{M_X}\bigg|^2
  -\frac{C^a_ib_ag^4_a(\mu)}{128\pi^4}|F_\phi|^2  \nonumber \\ 
  && \qquad\qquad\qquad +\frac{NC^a_ig_a^4(\mu)}{128\pi^4}
  \Big(\frac{F_X}{M_X}F_\phi^*+{\rm h.c.}\Big),
  \label{smgene}
\end{eqnarray}
where the summations for the gauge index $a$ are 
understood, and $C^a_i$ is the quadratic Casimir operator of gauge 
group $G_a$ for the field $Q_i$, for example, $(N_c^2-1)/2N_c$ for the
vectorial representation of $SU(N_c)$. For each formula, the first
term is the contribution of gauge threshold. This part determines the
mirage unification scale and the mirage mass spectrum as previously
shown for gaugino masses. The second term in each formula denotes the
anomaly mediation. The third term in the scalar mass 
squared $m_i^2$ is the mixed contribution of gauge and anomaly
mediations. It is noted that the relative complex phase 
of $F_X/M_X$ and $F_\Phi$ should be aligned from a phenomenological
viewpoint of CP violation. If this is the case, the mirage unification
does appear and further the third term in $m_i^2$ does not contain
CP-violating complex phases.

\bigskip\bigskip
{\centering\subsection{Mirage Unification}}

From the general formula \eqref{gmu_scale} in Section~\ref{sec:GMU},
the mirage unification scale for the present setup is found
\begin{eqnarray}
  \bar\mu_m \;=\; M_G\,\exp\bigg(\frac{-8\pi^2R}{Ng_G^2}\bigg)\,,
  \label{muMGM}
\end{eqnarray}
where $g_G^{}$ is the mirage value of unified gauge coupling
at $M_G$ and can be determined by evolving the low-energy observed
values up to high energy. Therefore, $g_G^{}$ and $M_G$ are
insensitive to the threshold scale and so is the mirage 
scale $\bar\mu_m$\@. It is noted that this is the general and
model-independent property of the theory with gauge coupling
unification such as the minimal supersymmetric SM\@. The real value of
unified gauge coupling, $g_U^{}$, in the present model is related 
to $g_G^{}$ as $\,1/g_U^2=1/g_G^2+(N/16\pi^2)\ln(M_X^2/M_G^2)$, 
but $g_U^{}$ itself does not appear explicitly in any formulas for
mirages. The parameter $R$ in \eqref{muMGM} is defined as the ratio of
two $F$ terms:
\begin{eqnarray}
  R \;=\; \frac{-F_\phi}{F_X/M_X},
\end{eqnarray}
which is real-valued as mentioned above and is defined so that its
sign becomes positive in most of known dynamics for 
the $X$ stabilization. In the limit $R\to0$ ($R\to\infty$), the
contribution of gauge (anomaly) mediation becomes dominant. It may be
interesting to see from Eq.~\eqref{muMGM} that the low-energy mirage
scale emerges as an analogy of dimensional transmutation: let us
consider a virtually-defined gauge coupling $g_m$. It has an initial
condition $g_m(M_G)=g_G^{}$ and obeys the renormalization-group
equation with the beta function coefficient $b_m=-N/R$ which is
negative in most cases, and hence $g_m$ has the asymptotically free
behavior.

Since the gauge couplings at $\bar\mu_m$ are related to $g_G^{}$ as
\begin{eqnarray}
  g_a^2(\bar\mu_m) \;=\; \frac{N}{N+b_aR}\,g_G^2\,,
  \label{gmum}
\end{eqnarray}
the soft supersymmetry-breaking mass parameters at the mirage
unification scale are found to be given by the following form:
\begin{eqnarray}
  M_{\lambda_a}(\bar\mu_m) \;&=&\; \frac{-Ng_G^2}{16\pi^2}
  \frac{F_X}{M_X}, \\[1mm]
  A_i(\bar\mu_m) \;&=&\; \frac{NC^a_i}{8\pi^2b_a}
  \big[g_G^2-g_a^2(M_X)\big]\frac{F_X}{M_X}, \\[1mm]
  m^2_i(\bar\mu_m) \;&=&\; 
  \frac{NC^a_i}{128\pi^4b_a}\Big[(N+b_a)g^4_a(M_X)-Ng_G^4\Big]
  \bigg|\frac{F_X}{M_X}\bigg|^2.
  \label{smMGM}
\end{eqnarray}
The mass spectrum is controlled by two parameters, the messenger
contribution $N$ and the threshold scale $M_X$, and is insensitive to
the $F$-term ratio $R$\@. On the other hand, the mirage 
scale $\bar\mu_m$ is determined by $N$ and $R$ and is insensitive to
the threshold scale $M_X$\@. These behaviors are important for
studying phenomenological aspects of the model, especially for
examining whether the mirage unification scale can be set to be
observable in future collider experiments, which we will discuss in
details in Section~\ref{sec:TeV}.

\bigskip\bigskip
{\centering\subsection{Mirage Spectrum}}

It is found from the above mass formula that the mirage gauge
mediation has a complete correspondence to the gauge mediation
scenario. That is, these two theories are traded to each other by
interchanging the gauge coupling constants: the mirage spectrum is
read off from the gauge-mediated one by simply replacing the gauge
couplings $g_a(\mu)$ at a low-energy scale $\mu$ with the mirage
unified value $g_G^{}$ which is evaluated from $g_a(\mu)$. Furthermore
soft scalar masses \eqref{smMGM} are found to generally satisfy two
types of sum rules, as in gauge mediation~\cite{MSS}: $\sum Ym^2=0$
and $\sum(B-L)m^2=0$ where $Y$ and $B-L$ are the hypercharge and the
baryon minus lepton number, respectively. (It is noted that each term
in \eqref{smgene}, i.e.\ the gauge, anomaly, mixed term, separately
satisfies the sum rules.)

The clear comparison to the gauge mediation model is summarized in
Table~\ref{tab:spectrum}. Here we show several illustrative limits 
that the mass spectra of two theories are evaluated at the same
low-energy scale ($=\text{TeV}$) in the cases that the
supersymmetry-breaking threshold scales are low ($M_X=\text{TeV}$) and
high ($M_X=M_G$).
\begin{table}[t]
\begin{center}
{\renewcommand{\arraystretch}{1.2}%
\begin{tabular}{lcc} \hline\hline
& ~~Low-scale mediation ~($M_X=\text{TeV}$)
& ~~~~High-scale mediation ~($M_X=M_G$)~ \\ \hline
\begin{tabular}{l}
GM\\ ($\mu=\text{TeV}$)
\end{tabular}
&
~\begin{tabular}{ccl}
$M_{\lambda_a}$ &=& $-Ng_a^2(\mu)$ \\
$A_i$ &=& $0$ \\
$m^2_i$ &=& $2NC^a_ig^4_a(\mu)$
\end{tabular}
& ~~~{\footnotesize%
\begin{tabular}{ccl}
$M_{\lambda_a}$ &=& $-Ng_a^2(\mu)$ \\[0.5mm]
$A_i$ &=& $\frac{2NC^a_i}{b_a}\big[g^2_a(\mu)-g_G^2\big]$ \\[0.5mm]
$m^2_i$ &=& 
$\frac{2NC^a_i}{b_a}\big[(b_a+N)g_G^4 -Ng_a^4(\mu)\big]$
\end{tabular}}
\\[8mm] \hline
\begin{tabular}{l}
Mirage GM\\ ($\bar\mu_m=\text{TeV}$)
\end{tabular}
& ~~~~~{\footnotesize%
\begin{tabular}{ccl}
$M_{\lambda_a}$ &=& $-Ng_G^2$ \\[0.5mm]
$A_i$ &=& $\frac{2NC^a_i}{b_a}\big[g_G^2-g^2_a(M_X)\big]$ \\[0.5mm]
$m^2_i$ &=& $\frac{2NC^a_i}{b_a}\big[(b_a+N)g^4_a(M_X)-Ng_G^4\big]$
\end{tabular}}
& 
~\begin{tabular}{ccl}
$M_{\lambda_a}$ &=& $-Ng_G^2$ \\
$A_i$ &=& $0$ \\
$m^2_i$ &=& $2NC^a_ig_G^4$
\end{tabular}
\\[8mm] \hline\hline
\end{tabular}
\bigskip
\caption{The soft supersymmetry-breaking parameters in the gauge and
mirage gauge mediations. In both cases, the parameters are evaluated
at the TeV scale. In this table, the gaugino masses $M_{\lambda_a}$
and trilinear couplings $A_i$ (scalar masses squared $m_i^2$) are
normalized by $F_X/16\pi^2M_X$ ($\,|F_X/16\pi^2M_X|^2$).\bigskip%
\label{tab:spectrum}}}
\end{center}
\end{table}
We have two typical spectra of the mirage unification:
\medskip
\begin{itemize}
\item The first is the
case that supersymmetry breaking is mediated at the gauge coupling
unification scale (the lower-right panel in the table). The low-energy
mass spectrum of gauginos is universal, the trilinear scalar couplings
vanish, and the scalar masses squared are specified by the quadratic
Casimir operators. The last fact means that scalar superparticles with
the same quantum charge have the universality of mass spectrum, which
leads to enough suppressions of rare processes involving
flavor-changing neutral currents. This virtue of the gauge mediation
also appears in the mirage gauge mediation. However, the mass spectrum
is rather different from the gauge mediation: the low-energy spectrum
is written only by the unified value of gauge couplings $g_G^{}$, not
the low-energy values. This fact leads to the low-energy unification
of gaugino masses as well as almost degenerate scalar
superparticles. The scalar lepton masses are of similar order of
scalar quark masses and they differ only 
by ${\cal O}(1)$ coefficients $C^a_i$. Moreover (too) restrictive mass
formulas generally imply several relations among observed mass values
in future collider experiments. For example, in the minimal
supersymmetric SM, the mirage spectrum is exactly given by
\begin{eqnarray}
  M_{\lambda_1}^2:M_{\lambda_2}^2:M_{\lambda_3}^2:
  m_Q^2:m_u^2:m_d^2:m_L^2:m_e^2 \,=\, 
  N:N:N:\frac{21}{5}:\frac{16}{5}:\frac{14}{5}:
  \frac{9}{5}:\frac{6}{5},\qquad
\end{eqnarray}
without including Yukawa coupling effects. Therefore the whole
superpartners are found to receive a similar size of
supersymmetry-breaking masses.

\item The second limit is the low-scale threshold (the lower-left
panel in the table). Here we discuss the 
situation $M_X\sim\bar\mu_m\sim\text{TeV}$\@. The low-energy gaugino
masses are universal and given by $g_G^{}$, which is a robust
prediction of the mirage gauge mediation. Unlike the usual
(low-scale) gauge mediation, scalar trilinear couplings are generated
at one-loop order of gauge couplings and naturally comparable to other
supersymmetry-breaking parameters. For 
example, if $M_X=\bar\mu_m$, the trilinear couplings are found from
the above formula and \eqref{gmum} 
that $A_i=\frac{g_G^2}{8\pi^2}\frac{NRC^a_i}{N+b_aR}
\frac{F_X}{M_X}\,\sim M_\lambda$. Such sizable $A$ parameters would be
important for phenomenology around the electroweak symmetry breaking
scale. The scalar soft mass parameters $m_i^2$ are also
characteristic. In the gauge mediation with chiral 
messengers, $m_i^2$ is positive irrespectively of the threshold 
scale $M_X$, but the low-scale mirage gauge mediation sometimes
predicts tachyonic scalar superpartners. For 
example, if $M_X=\bar\mu_m$, we find from the mass formula and the
gauge coupling relation \eqref{gmum} that the positivity constraint of
scalar masses squared ($m_i^2>0$) lead to an inequality
\begin{eqnarray}
  b_aR^2\;<\;N(1-2R)
  \label{ineq}
\end{eqnarray}
for all beta function coefficients $b_a$. That implies 
that, for $R>1$ ($R>1/2$), asymptotically free (non-free) gauge groups
induce tachyonic contributions to scalar soft masses. It is therefore
important for phenomenology of the model to satisfy some lower bound
on the threshold scale $M_X$ and/or an upper bound on the 
ratio $R$, the latter of which restricts possible dynamics for 
the $X$ stabilization.
\end{itemize}

\bigskip\bigskip\bigskip
{\centering\section{Mirage Unification at TeV}
\label{sec:TeV}}

Based on the formalism shown above, we investigate the model with
mirage gauge mediation around the TeV 
scale, i.e.\ $\bar\mu_m\sim\text{TeV}$\@. The forthcoming Large Hadron
Collider experiment will probe the TeV-scale physics, and in
particular, would observe the superpartners of SM fields with the
mirage pattern of mass spectrum. Such a characteristic spectrum
provides distinctive experimental signatures from any other
supersymmetry-breaking scenarios, and clearly suggests the existence
of some specific mechanism in high-scale dynamics. In this section, we
first derive the conditions for realizing TeV-scale unification. Next,
we examine several models of $X$ field dynamics discussed in the
literature and show that it seems difficult for these models to
satisfy the required conditions. Finally, possible dynamical
mechanisms are presented to make the conditions unnecessary or
weakened. We also point out that the hidden sector contribution, which
is generally needed to have the de Sitter vacuum in supergravity but
is usually decoupled, may play an important role for constructing a
full theory of TeV mirage unification.

\bigskip\bigskip
{\centering\subsection{TeV-scale Mirage}
\label{sec:TeVMGM}}

For phenomenological discussions of mirage gauge mediation, there are
three points to be taken into account: (i) the perturbative evolution
of gauge coupling constants, (ii) non-tachyonic scalar mass spectrum,
and (iii) the low mirage scale.

As for the first point, the one-loop evolution of gauge couplings are
solved as
\begin{eqnarray}
  \frac{1}{g_U^2} \;=\; \frac{1}{g_a^2(\mu)}
  +\frac{b_a}{16\pi^2}\ln\left(\frac{\mu^2}{M_X^2}\right)
  +\frac{b_a+N}{16\pi^2}\ln\left(\frac{M_X^2}{M_G^2}\right),
\end{eqnarray}
for $\mu<M_X<M_G$. The unification scale is determined by low-energy
observed values $g_a^2(\bar\mu_m)$ and the requirement of gauge
coupling unification, independently of other parameters. Therefore the
running of gauge couplings, in particular, their high-energy values
are controlled by the threshold scale $M_X$ and the number of
messenger fields $N$. A bound on these parameters is derived from the
requirement of perturbative unification that the gauge couplings do
not diverge below the unification scale (i.e.\ $g_U^{}<\infty$):
\begin{eqnarray}
  N\,\ln\left(\frac{M_G}{M_X}\right) \;<\; \frac{8\pi^2}{g_G^2}\,.
  \label{perturb}
\end{eqnarray}
This inequality implies the lower bound on the messenger 
mass $M_X$ and the upper bound on its number $N$\@. For example, we
obtain from
\eqref{perturb}
\begin{eqnarray}
  N\,=\,\;\,5 \;&:&\;\; M_X\,>\;4.5\times10^2~~\text{GeV}, \\
  N\,=\,10 \;&:&\;\; M_X\,>\;3.0\times10^9~~\text{GeV}, \\
  N\,=\,15 \;&:&\;\; M_X\,>\;5.7\times10^{11}~\text{GeV},
\end{eqnarray}
for $M_G=2.0\times10^{16}$ GeV which is a typical scale of
supersymmetric grand unification of the SM gauge couplings.

\begin{figure}
\begin{center}
\includegraphics[width=10cm,clip]{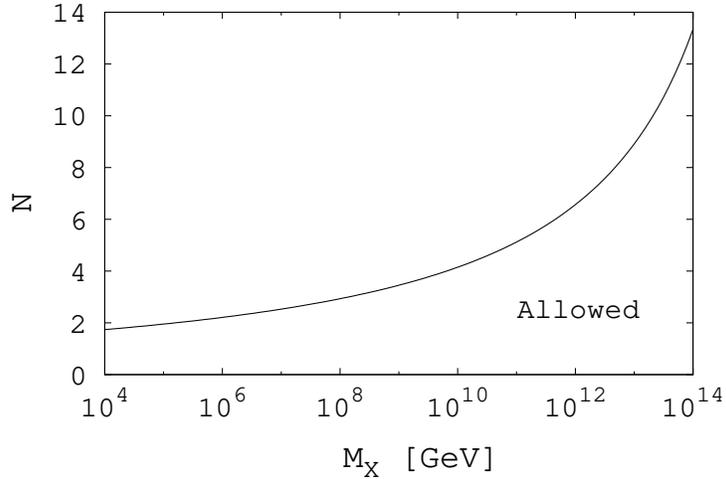}
\caption{The region for the messenger parameters $M_X$ and $N$ allowed
by the positivity constraint on the right-handed scalar lepton mass in
the minimal supersymmetric SM\@.\bigskip%
\label{fig:me}}
\end{center}
\end{figure}

The second condition comes from the superparticle mass spectrum at a
low-energy observable scale. In order that charged scalar
superpartners do not develop condensations, their mass-squared terms
in the potential must be positive. Here we consider the constraint
that soft supersymmetry-breaking masses squared $m_i^2$ must be
positive, as a conservative one without including the effects of
Yukawa couplings and trilinear scalar parameters. The analysis in the
previous section shows that the scalar masses squared become at the
mirage scale
\begin{eqnarray}
  m^2_i(\bar\mu_m) \;=\; 
  \frac{NC^a_i}{128\pi^4b_a}\Big[(N+b_a)g^4_a(M_X)
  -Ng_G^4\Big]\bigg|\frac{F_X}{M_X}\bigg|^2.
  \label{scalarm2}
\end{eqnarray}
The gauge couplings at the intermediate scale, $g_a(M_X)$, are
determined by $M_X$ for fixed values of low-energy gauge
couplings. Therefore the scalar masses are controlled by the two 
parameters $M_X$ and $N$\@. Roughly speaking, tachyonic scalars are
avoided if $m^2_i(\bar\mu_m)>0$, namely, the quantity in the bracket
of \eqref{scalarm2} is negative (positive) for asymptotically free
(non-free) gauge theory. For example, in the minimal supersymmetric
SM, the right-handed scalar leptons usually give the most significant
constraint. We 
find from \eqref{scalarm2} that $m^2_e(\bar\mu_m)>0$ implies
\begin{eqnarray}
  b_1R(M_X)^2 \;<\; N\big[1-2R(M_X)\big],
\end{eqnarray}
where $b_1=33/5$ is the beta function coefficient for the hypercharge
gauge coupling, and $R(M_X)$ has been introduced as a generalization
of \eqref{ineq} and defined 
as $R(M_X)\equiv(Ng_G^2/8\pi^2)\ln(M_G/M_X)$\@. It is easily found
that the number of messengers $N$ has an upper bound for their masses
fixed, and in other words, the threshold scale $M_X$ has a lower
bound. In Fig.~\ref{fig:me}, we show the numerical result of the
positivity constraint $m^2_e(\bar\mu_m)>0$.
The parameter bounds are often more severe than \eqref{perturb} which
is obtained from the perturbative gauge coupling unification.

The last point is whether the mirage unification takes place at a
low-energy observable scale as one chooses. The low-energy unification
scale in the mirage gauge mediation is found in the previous analysis
[Eq.~\eqref{muMGM}]: 
\begin{eqnarray}
  \bar\mu_m \;=\; M_G\,\exp\bigg(\frac{-8\pi^2R}{Ng_G^2}\bigg)\,.
\end{eqnarray}
Since $M_G$ and $g_G^{}$ are determined by low-energy theory and
observations, the scale $\bar\mu_m$ is controlled 
by $N$ and $R$\@. The latter is defined by the ratio of two
supersymmetry-breaking $F$ terms in the theory and then, possible
dynamics of $X$ is restricted for the TeV-scale mirage to be achieved:
\begin{eqnarray}
  \frac{R}{N} \;=\; \frac{g_G^2}{8\pi^2}
  \ln\left(\frac{M_G}{\text{TeV}}\right) \;\simeq\; 0.20\,.
  \label{RN}
\end{eqnarray}
Here we have used a typical unified value of SM gauge 
couplings, $g_G^2/4\pi^2=1/24.5$, which is obtained from the
weak-scale experimental data and the renormalization-group evolution
in the minimal supersymmetric SM~\cite{GUT}. The result \eqref{RN} is
deeply related to high-energy dynamics of supersymmetry breaking. As a
simple example, let us consider the case that the K\"ahler potential
is minimal and the superpotential contains a single polynomial
term~\cite{DAM}:
\begin{eqnarray}
  K \;=\; Z|X|^2, \qquad\quad
  W \;=\; X^n \quad (n\geq3).
\end{eqnarray}
Turning on the compensator $F$ term as a background in the
Lagrangian\footnote{The compensator $F_\phi$ is dynamically 
fixed, e.g.\ by including a constant ($X$-independent) superpotential
term in supergravity. Such a detail is irrelevant to the result
presented here.} and minimizing the supergravity scalar potential, one
obtains a non-vanishing $F$ component of supersymmetry-breaking 
field $X$:
\begin{eqnarray}
  \frac{F_X}{X} \;=\;\frac{2}{1-n}F_\phi\,, \qquad
  \Big(\,R=\frac{n-1}{2}\,\Big).
\end{eqnarray}
In order to satisfy \eqref{RN}, the correlation is found between the
potential form ($W=X^n$) and the number of messenger fields, as given
in the following:
\begin{eqnarray}
\begin{array}{c|ccccccc}\hline\hline
~n~~ & 3 & 4 & 5 & 6 & 7 & 8 & 9 \\ \hline
~N~~ & ~5.0~ & ~7.5~ & ~10.0~ & ~12.6~ & ~15.1~ & 
~17.6~ & ~20.1~ \\ \hline\hline
\end{array}
\end{eqnarray}
An integer value of the messenger number may be reasonable. It is
interesting that the simplest dynamics, $W=X^3$, is approximately
consistent with the TeV-scale mirage for an integer number $N=5$. A
variety of other models have been proposed in the literature to
stabilize $X$ with the $F_\Phi$ effect and to provide
non-supersymmetric (non-decoupling) messenger thresholds. The
potentials and their predictions are summarized 
in Table~\ref{tab:models}, in which we also show the $X$ scalar 
mass $m_X$ in each supersymmetry-breaking vacuum.
\begin{table}[t]
\begin{center}
\begin{tabular}{lcccc} \hline\hline
\multicolumn{1}{c}{dynamics} & $R$ & $M_X$ & $m_X$ & \\ \hline
\begin{tabular}{l}
$K=|X|^2$\\
$W=X^n\;\;\;(n>3)$
\end{tabular}
& $\dfrac{n-1}{2}$
& $F_\phi^\frac{1}{n-2}$
& $(n-3)|F_\phi|$ 
& ~\cite{DAM}
\\ \hline
\begin{tabular}{l}
$K=Z(X,X^\dagger)|X|^2$\\
$W=X^3$
\end{tabular}
& $1$ 
& arbitrary
& $\text{(2-loop)}\times|F_\phi|$
& ~\cite{DAM}
\\ \hline
\begin{tabular}{l}
$K=|X|^2+|Y|^2$\\
$W=X^nY^m$
\end{tabular}
& ~{\footnotesize$\dfrac{m+n-1}{2}$}~
& $F_\phi^\frac{1}{m+n-2}$
& ~$(m+n-3)|F_\phi|$~
& ~\cite{axion}
\\ \hline
\begin{tabular}{l}
$K=Z(X,X^\dagger)|X|^2-|c||X|^4$~~\\
$W=0$
\end{tabular}
& $1$ 
& ~~$\text{(1-loop)}\times\Lambda$~~
& $\text{(2-loop)}\times|F_\phi|$
& ~\cite{axion}
\\ \hline
\begin{tabular}{l}
$\,$no $X$\\
$\Delta K=\Psi\bar\Psi$\end{tabular}
& $\dfrac{1}{2}$ & ${\cal O}(F_\phi)$ & $-$
& ~\cite{NW}
\\ \hline\hline
\end{tabular}
\bigskip
\caption{Typical dynamics for supersymmetry-breaking messenger mass
splitting. The limit $R\to\infty$ corresponds to the anomaly mediation
dominant (supersymmetric thresholds), and $M_X$ and $m_X$ denote the
threshold scale and the $X$ scalar mass in the vacuum,
respectively.\bigskip%
\label{tab:models}}
\end{center}
\end{table}
It is found from the table that, in most of models, the $F$-term 
ratio $R$ is ${\cal O}(1)$ or sometimes becomes large. This fact
generally means that the number of messenger fields is required to be
large for the TeV-scale mirage. If this is the case, the messenger
mass scale should be unfortunately high in order to have the
perturbative gauge coupling unification or not to have any tachyonic
scalar superpartners.

\bigskip\bigskip
{\centering\subsection{Possible Ways Out}}

We have studied the phenomenological constraints in the simple case of
mirage gauge mediation and found that it tends to need a large number
of messenger fields and a high mediation scale. It is a natural
amelioration to realize a low-scale mirage unification without
introducing model complexity and/or without loosing observation
feasibility.

If the messenger number $N$ becomes large, supersymmetry-breaking
masses squared of scalar superpartners become negative, as seen in the
previous section. It is then possible to introduce some additional
dynamics for stabilizing these tachyons in parallel ways to various
proposed solutions of the tachyonic scalar lepton problem in the pure
anomaly mediation~\cite{AM_RS}. It may be interesting to look for
tachyon-stabilization dynamics which is characteristic to the mirage
gauge mediation.

Another remedy is found from the expression of the mirage 
scale \eqref{muMGM} that if the virtual unified gauge 
coupling $g_G^{}$ is increased, the messenger number $N$ can be
correspondingly reduced for a fixed value of mirage scale. A
high-energy gauge coupling is generally increased by introducing
additional fields. It is however noted that, as shown in
Section~\ref{sec:c1}, the virtual gauge coupling $g_G^{}$ is
insensitive to the existence of intermediate-scale thresholds and is
fixed only by low-energy physics. Therefore one is lead to modifying
low-energy physics by adding TeV-scale extra fields. We suppose that
these fields belong to grand unification-like and vector-like
representations in order to preserve the gauge coupling unification
and to avoid the experimental constraints from precision
electroweak-scale measurements. The extra fields are assumed to be
irrelevant to supersymmetry breaking and their threshold is
supersymmetric. In this case, $g_G^{}$ is increased as
\begin{eqnarray}
  \frac{g_G'^{\,2}}{g_G^2\,} \;=\; 
  \frac{1}{1-\frac{\Delta b\,g_G^2}{8\pi^2}
    \ln\big(\frac{M_G}{\text{TeV}}\big)}\,,
\end{eqnarray}
where $\Delta b$ denotes the universal extra-field contribution to 
beta function coefficients ($\Delta b>0$). We find that the number of
supersymmetry-breaking messengers is reduced for a fixed mirage scale:
\begin{eqnarray}
  \frac{N'}{R'} \;=\; \frac{N}{R}-\Delta b.
  \label{RN2}
\end{eqnarray}
The ratio of the messenger number and the $F$-term ratio is determined
by low-energy physics, and $N/R\simeq5.0$ in the minimal
supersymmetric SM [Eq.~\eqref{RN}]. As a simple example, if we add one
pair of $16$ and $16^*$ representations of $SO(10)$ at the TeV
scale~\cite{extra}, $\Delta b=4$ and hence the minimal 
messenger ($N'=1$ and $R'=1$) is sufficient to obtain the mirage
phenomenon, where tachyonic scalar superpartners do not emerge (see
Fig.~\ref{fig:me}).

A more reasonable solution is to reduce $R$ in a dynamical way. It is
found from \eqref{RN} that a smaller (positive) $R$ implies a fewer
messenger multiplets needed and the model becomes simplified. For
example, if we have some dynamics which predicts $R\simeq1/5$, only a
single pair of messengers is sufficient to realize the TeV-scale
mirage unification. Since a smaller value of $|R|$ means a larger
effect of $F_X$ relative to $F_\phi$, some mechanism of the $X$ field
is needed to amplify its supersymmetry-breaking effect a few times or
so.
\medskip
\begin{itemize}
\setlength{\itemsep}{7pt}
\setlength{\parskip}{0pt}
\setlength{\parindent}{11pt}
\item Multiple $X$ fields :\\[1mm]
One may naively expect that the threshold contribution increases when
several supersymmetry-breaking fields are introduced with
non-vanishing $F$ components. However the total effect of
supersymmetry breaking is not enhanced if these $X$ fields have
similar types of dynamics and then induce similar orders of $F$ terms:
the resulting effect from multiple $X$ fields is not additive and is
the same as the single $X$ case. This behavior is confirmed for
various types of $X$ dynamics (e.g.\ see \cite{axion}). In the end, a
viable model along this line must be constructed to have highly
asymmetric property among multiple $X$ fields. That generally makes
the model complex and unrealistic.

\item Different $X$ potentials :\\[1mm]
In the above example of mirage gauge mediation, the K\"ahler and
superpotential of $X$ are minimal and simplest. A model with different
type of $X$ potential may lead to increasing the
supersymmetry-breaking effect $|F_X/X|$ and then reducing $R$\@. As we
will show in details, the supergravity analysis of $F$ terms leads to
the following form of the $R$ parameter in the vacuum
\begin{eqnarray}
  R \;=\; \frac{W_{XX}+\frac{4}{3}X^\dagger W_X+
    \frac{1}{3}X^{\dagger2}W}{2W_X/X},
\end{eqnarray}
for the minimal K\"ahler $K=|X|^2$ and general 
superpotential $W(X)$. This expression has been written down by
neglecting higher-order terms in $X$ and without including the hidden
sector effect, for simplicity. The exploration 
of $W(X)$ realizing $R\simeq0.2N\lesssim1$ is an interesting task to
be performed. We will later discuss it in several examples including
the hidden sector contribution. The $R$ parameter is sometimes
determined by continuous model parameters. In this case, the mirage
scale is set just by choosing these parameters non-dynamically, while
it is preferable that the mirage is described in terms of discrete
parameters which define the dynamics of model such as the power of
polynomial potential.

\item Different messenger couplings :\\[1mm]
The messenger supermultiplets $\Psi$ and $\bar\Psi$ are coupled 
to $X$ and the compensator $\Phi$ somewhere in the Lagrangian and
receive supersymmetry-breaking mass splitting within each multiplet
when the $F$ components $F_X$ and $F_\Phi$ are turned on.

A simple and direct way to modify the mass splitting is to introduce
extra quadratic terms in the K\"ahler and superpotential:
\begin{eqnarray}
  \Delta K \;=\; \bar\Psi\Psi, \qquad\qquad
  \Delta W \;=\; M_\Psi\bar\Psi\Psi,
\end{eqnarray}
in addition to the basic Lagrangian of mirage gauge 
mediation \eqref{LagMGM}. The messenger scalars receive additional
supersymmetry-breaking masses induced from these terms as well as the
supersymmetric mass $M_\Psi$. The modification of the model is easily
found by noticing that the inclusion of additional supersymmetry
breaking is effectively described by the field 
redefinition: $X\to X+M_\Psi+F_\phi^*/\Phi^2$. Therefore the 
modified $R$ parameter is read off as
\begin{eqnarray}
  R \;=\; \frac{(X+M_\Psi)F_\phi+|F_\phi|^2}{2|F_\phi|^2-F_X}\,.
\end{eqnarray}
The new contribution becomes significant only when the messenger mass
scale is low: $M_X\sim M_\Psi\sim F_\phi$. If this application limit
is acceptable, the TeV-scale mirage may be realized with a fewer
number of messengers by taking appropriate values of 
the $\Delta K$ and $\Delta W$ contributions. For example, if 
the $\Delta K$ effect is dominant, $R$ is reduced to $1/2$ and 
Eq.~\eqref{RN} requires $N\simeq2.5$, which is the half of the
previous result in the simplest case ($R=1$).

The messenger coupling to $X$ is another possible source of modifying
the supersymmetry-breaking mass splitting and reducing the number of
messengers. Let us consider the following form of messenger coupling
\begin{eqnarray}
  W \;=\; X^m\bar\Psi\Psi.
\end{eqnarray}
This form can be general by assigning suitable $R$ symmetry
charges. The $m=1$ case is simplest and has been analyzed before. The 
superpotential coupling determines the messenger mass scale as well as
the strength of supersymmetry-breaking 
mediation. Inserting $X=M_X+F_X\theta^2$, we find that the effective
messenger mass $M_{\rm eff}$ and supersymmetry-breaking mass 
splitting $F_{X_{\rm eff}}$ are given by
\begin{eqnarray}
  M_{\rm eff} \;=\; (M_X)^m, \qquad\qquad
  \frac{F_{X_{\rm eff}}}{M_{\rm eff}} \;=\; m\frac{F_X}{M_X}.
\end{eqnarray}
The supersymmetry-breaking effect is thus enhanced by the 
factor $m$ compared with the usual $m=1$ case and so the $R$ parameter
is reduced by the same factor. In the end, the number of messengers can
be reduced. One price to pay is that the messenger mass scale is no
longer free and is suppressed from the mediation 
scale $M_X$ as $M_{\rm eff}=
(\frac{M_X}{\Lambda})^{m-1}M_X$ where $\Lambda$ is the ultraviolet
cutoff. It should be noted that the mirage unification and its
emergence scale are not affected by changing the mass scales of
messenger fields and supersymmetry breaking, as we have shown. A
phenomenological bound on the messenger mass 
scale, i.e.\ $M_{\rm eff}>\text{TeV}$, leads to a restriction of
messenger coupling, in particular, an upper bound on the index $m$ as
a function of the supersymmetry-breaking scale $M_X$. For example, if
one takes $\Lambda=M_{\rm pl}$ and $M_X=M_G$, the index must 
satisfy $m<7.6$. Therefore a favorable value, $m=5$, for the TeV-scale
mirage [see \eqref{RN}] is within the allowed range. In other words,
the superpotential coupling with $m=5$ generates the messenger mass
around $100$~PeV scale, enough high to satisfy experimental constraints.

\item Extra sources of supersymmetry breaking :\\[1mm]
The enhancement of $F_X$ effect is effectively done by introducing
extra sources of supersymmetry breaking other than the $X$ field. We
here comment on several possibilities in order.

It is a natural expectation that a better way to modify a model
involves fewer extensions of it. In this sense, a simple way is to
consider the complete anomaly mediation, i.e.\ to include the
supersymmetry-breaking effects induced not only from the super-Weyl
anomaly but also from other anomalies in supergravity. The latter
effects may be comparable in some framework to that of the conformal
compensator and then our previous results for the $R$ parameter may be
changed.

Extra supersymmetry-breaking effects are supposed to have the property
that the mirage unification is not disturbed. The general form of such
supersymmetry breaking is parameterized as \eqref{sol}. That is, in
addition to the gauge threshold effect (the $c_1$ term) analyzed
before, some universal contribution (the $c_0$ term) can be
included. If these two contributions are on similar orders of magnitude
(and have the same sign), the supersymmetry-breaking effect is
effectively enhanced and the number of messengers may be reduced to a
reasonable level. A plausible possibility of the universal
contribution comes from the gravity and related modulus fields, which
have field-universal interactions. A well-known framework of moduli
stabilization in string theory~\cite{KKLT} provides such a
possibility~\cite{XT}. It is important to notice that in this
framework the modulus contribution to supersymmetry breaking is found
to be comparable to the anomaly mediated one~\cite{CFNO,CJO,EYY} and
hence also comparable to the gauge threshold contribution. This is the
property we just wanted in the above for suitably improving the mirage
gauge mediation.
\end{itemize}

\bigskip\bigskip
{\centering\subsection{Hidden Sector and Mirage Gauge Mediation}}

The above analysis has not been concerned about the vacuum energy (the
cosmological constant). At the minimum of potential, the scalar
component of the $X$ field (related to the messenger mass scale) is
taken to be suppressed and then the vacuum energy is negative. We
therefore need to uplift the potential to make the cosmological
constant zero or slightly positive. This point has not been discarded
in the literature of deflected anomaly mediation or simply regarded as
adding the hidden sector which decouples from $X$\@. In this section,
we examine the possibility that the mirage gauge mediation, in
particular the $R$ parameter, is modified by utilizing hidden-sector
dynamics for uplifting the vacuum energy. It is better to realize that
the modification is done such that the TeV-scale mirage naturally
emerges in a simpler model and the mirage scale is controlled by
discrete parameters. It has been known~\cite{PT} that the TeV-scale
unification in the simple mirage case~\cite{TMM} is difficult to
realize. It would be therefore interesting that the mirage gauge
mediation solves this problem with the hidden sector uplifting which
is experimentally required for the cosmological observation.

\bigskip\bigskip
{\centering\subsubsection{Hidden Sector Contribution}}

We introduce a hidden-sector field $Z$ with a non-vanishing vacuum
expectation value of $F$ component. The general supergravity
Lagrangian for $Z$ and the supersymmetry-breaking field $X$ has the
following form:
\begin{eqnarray}
  {\cal L}_H \;=\; \int\!d^4\theta\,\Phi^\dagger\Phi\,
  f(X,X^\dagger\!,Z,Z^\dagger) \,+\Big[\int\!d^2\theta\,\Phi^3\,
  W(X,Z) \,+{\rm h.c.}\Big].
\end{eqnarray}
The supergravity $f$ function is related to the K\"ahler potential 
as $f=-3e^{-K/3}$. The loop-level dependence on the 
compensator $\Phi$ has been dropped since it is quantitatively
irrelevant to the discussion in this section. One can incorporate 
in $f$ the direct couplings between $X$ and $Z$ without conflicting
with phenomenological observation in the visible sector. Integrating
out the hidden sector, we obtain the supergravity scalar potential
\begin{eqnarray}
  V_H \;=\; e^{K/3}\Big[(WK_X+W_X)K^{-1}_{XX^\dagger}
  (W^*K_{X^\dagger}+W^*_{X^\dagger})-3|W|^2\Big] 
  +f_{ZZ^\dagger}|F_Z|^2,
\end{eqnarray}
where the lower indices of $f$ and $W$ denote the field
derivatives. We consider that the potential $V_H$ is a function of 
the $X$ field, and the hidden variable is treated as a background
parameter which is determined by solving the $Z$ dynamics in the
hidden sector.

In this paper we explore the mirage gauge mediation which has the
parameter region: $|F_X/X|\sim|F_\phi|$ for the mirage to appear 
and $|X|\ll1$ for the messengers to be lighter than the cutoff
scale. It is noted that, in a complete contrast, $|F_X/X|\ll|F_\phi|$
and $|X|\sim1$ in the scenario of string-theory moduli
stabilization. Therefore the framework of supersymmetry-breaking and
uplifting hidden dynamics is expected to be different from the
string-theory scenario~\cite{uplift}. The smallness of expectation 
value $|X|\ll1$ may naturally lead to the conditions for the K\"ahler
potential that $|XK_X|\ll1$ and $|XX^\dagger K_{XX^\dagger}|\ll1$ at
the minimum. If this is the case, the vacuum energy is easily found
\begin{eqnarray}
  V_{H0} \;=\; -3e^{-K/3}|F_\phi|^2+f_{ZZ^\dagger}|F_Z|^2.
\end{eqnarray}
The requirement of the vanishing cosmological constant is fulfilled
with a non-vanishing $F$ term of hidden-sector field $Z$ in the
vacuum. Minimizing the potential $V_H$ with respect to the $X$ scalar,
we find the shifted vacuum by turning on $F_Z$. Substituting these
results, we obtain the $F$ components $F_X$ and $F_\Phi$ in the
shifted vacuum, in particular, the general formula of their ratio in
the leading order:
\begin{eqnarray}
  R \;=\; \frac{W_{XX}+W_XK_X+WK_{XX}+(W_X+WK_X)\big[\frac{1}{3}K_X
    +\big(K^{-1}_{XX^\dagger}\big)_XK_{XX^\dagger}\big]\,}{
    2W_X/X-3Wf_{XZZ^\dagger}/Xf_{ZZ^\dagger}}\,.
\end{eqnarray}
For example, the $R$ parameter is evaluated for the minimal form of
K\"ahler potential $K=|X|^2+|Z|^2$ as
\begin{eqnarray}
  R_{\rm min} \;=\; \frac{W_{XX}+\frac{4}{3}X^\dagger W_X
    +\frac{1}{3}X^{\dagger2}W}{2W_X/X+X^\dagger W/X}\,.
\end{eqnarray}
Since the second term in the denominator expresses the hidden sector
contribution, the uplifting of vacuum energy is found to multiply 
the $F_X$ effect by the factor $H$:
\begin{eqnarray}
  H \;\equiv\; 1+\frac{X^\dagger W}{2W_X}.
  \label{H}
\end{eqnarray}
This formula of the enhancement is given only by the superpotential
for the $X$ field. If the dynamics for $X$ stabilization 
satisfies $H>1$, the hidden sector enhances the $F_X$ effect which
implies that the number of messenger fields is effectively reduced and
tachyonic scalar mass spectrum is avoided. Moreover 
the $H$ factor \eqref{H} indicates that the ratio of two $F$ terms
remains real and does not disturb the phase alignment of
supersymmetry-breaking soft mass parameters.

\bigskip\bigskip
{\centering\subsubsection{Sample Potentials}}

In this subsection, we assume that the K\"ahler potential has the
minimal form: $K=|X|^2+|Z|^2$ as the simplest case, and examine
several forms of superpotential for $X$ to have a suitable value of
the enhancement factor $H$.
\medskip
\begin{itemize}
\setlength{\itemsep}{7pt}
\setlength{\parskip}{0pt}
\setlength{\parindent}{11pt}
\item $W=yX^n+c$ ~~($n>3$) :\\[1mm]
The first example is the polynomial superpotential discussed in
Section~\ref{sec:TeVMGM}. Here we also include a constant
superpotential term to dynamically stabilize $F_\phi$. The analysis of
supergravity potential is found to give the minimum at
\begin{eqnarray}
  \frac{X^n}{|X|^2} \;=\; \frac{3-n}{n(n-1)}\,\frac{c}{y}\,,
\end{eqnarray}
for $|y|\gg|c|$. From \eqref{H}, we obtain the factor $H$ as
\begin{eqnarray}
  H \;=\; \frac{n-5}{2n-6}\,.
\end{eqnarray}
While $H$ becomes a real parameter, it generally 
takes $H<\frac{1}{2}$ and cannot be used to effectively enhance 
the $F_X$ effect.

\item $W=yX+c$ :\\[1mm]
The second is the linear superpotential term (so to say, a low-scale
Polonyi model). This model has a different type of minimum than the
above polynomial superpotential with a higher 
power. For $|y|\ll|c|$, the supergravity potential is minimized at
\begin{eqnarray}
  X^3 \;=\; \frac{6cy^{*2}}{c^{*2}y}\,.
\end{eqnarray}
The model predicts $R=1$ without taking into account the uplifting
hidden sector. The hidden-sector enhancement factor is given by
\begin{eqnarray}
  H \;=\; 1+\frac{cX^\dagger}{2y} \;\simeq\; 
  \left(\frac{|c|}{|y|}\right)^{2/3}.
\end{eqnarray}
Since $H$ becomes large and positive, the $F_X$ effect is enhanced in
the uplifted true vacuum. So the TeV-scale mirage can be made
natural. It is however noted that the $F$-term ratio $R$ is no longer
a discrete value and depends on the continuous coupling constants of
the model.

\item $W=y_nX^n+y_mX^m$ :\\[1mm]
The third model is the racetrack-like superpotential. That is, the two
similar superpotential terms work in cooperation to stabilize 
the $X$ field. Analyzing the supergravity potential, we obtain the
minimum at
\begin{eqnarray}
  X^{m-n} \;=\; -\,\frac{ny_n}{my_m},
  \label{Xracet}
\end{eqnarray}
where the relative size of $y_m$ and $y_n$ is assumed to 
have $|X|\ll1$. The model predicts $R=1$ without the hidden sector
contribution. It is noticed that, with this expectation value 
of $X$ \eqref{Xracet}, the first derivative of the superpotential
vanishes and the previous formula \eqref{H} cannot be used. In this
case, the general (re)analysis of potential minimization and the
vacuum energy uplifting lead to
\begin{eqnarray}
  R \;=\; \frac{1}{1+ff_{XZZ^\dagger}W/f_{ZZ^\dagger}XW_{XX}}.
\end{eqnarray}
From this formula, the $H$ factor is evaluated for the minimal
K\"ahler potential:
\begin{eqnarray}
  H \;=\; 1+\frac{X^\dagger W}{XW_{XX}} \;=\; 1-\frac{|X|^2}{mn}.
\end{eqnarray}
In the end, we find $H\simeq1$ and the hidden-sector effect is
negligible in this model.

\end{itemize}
We have investigated three types of models and found three different
conclusions. All of these models unfortunately have somewhat
unsatisfied points. The improvement and construction of realistic
models are left for future study.

\bigskip\bigskip\bigskip
{\centering\section{Summary and Discussion}}

In this paper, we investigated a new class of supersymmetry-breaking
mediation models, where gaugino masses are unified in the low-energy
regime. We first classified the conditions of gaugino mass
unification, and then studied the gauge threshold case. The mirage
gauge mediation scenario is basically the gauge mediated supersymmetry
breaking, but at the low-energy unification scale, the virtual
high-energy unified gauge coupling behaves as the one at the 
renormalization scale in gauge mediation. Thus under the hypothesis of
gauge coupling unification, gaugino masses become naturally unified at
the weak scale. On the other hand, it is non-trivial to dynamically
realize the mirage unification at the TeV scale. We also discussed
several possible ways out in the last part of this paper.

The mirage gauge mediation possesses the characteristic mass spectrum
of superparticles and various virtues from the phenomenological points
of view. Compared with the gauge mediation, the masses of
superparticles tend to be degenerate at the weak scale. Also the
gravitino is not the lightest superparticle anymore, but is rather
heavy to have sizable corrections from the anomaly mediation. On the
other hand, unlike the simple mirage case such as the string-theory
framework of moduli stabilization, the low-scale gaugino mass
unification is not an assumption but is a natural prediction of the
mirage gauge mediation. In addition, thanks to the virtues of gauge
mediation, the flavor-changing rare processes and CP violations are
automatically suppressed.

The mirage gauge mediation is favored as well from the cosmological
points of view. The scenario contains a singlet scalar 
field, $X$, coupled to supersymmetry-breaking messenger 
fields. Since $X$ has a rather flat potential, it is considered to
dominate the energy of the Universe. Then the $X$ scalar decays into
superparticles and gravitinos, diluting the pre-existing particles
and producing radiations. The produced gravitinos often easily spoil
the successes of the standard cosmology such as the big-bang  
nucleosynthesis, or overclose the Universe. However it is expected in
our scenario that the branching ratio of the gravitino production
becomes suppressed since the vacuum expectation value of $X$ is much
smaller than the Planck scale (see e.g.~\cite{gravitinosup}). This
feature is contrasted to the string-related mirage models, where a
light modulus field is involved and causes a serious problem of the
gravitino overproduction~\cite{m-induced}. The dark matter candidates
in the mirage gauge mediation are the (degenerate) gauginos and the
superpartner of $X$\@. They are produced from the decay of $X$ scalar,
while their relic abundance is quite model dependent. We need further
studies on the phenomenological and cosmological aspects of the
scenario, and they will be discussed in the future.

\bigskip\bigskip\bigskip
{\centering\acknowledgments}

This work was supported in part by the grant-in-aid for scientific 
research on the priority area (\#441) ``Progress in elementary
particle physics of the 21st century through discoveries of Higgs
boson and supersymmetry'' (No.~16081209) and by a scientific grant 
from the ministry of education, science, sports, and culture of Japan
(No.~17740150).

\newpage

\end{document}